\documentclass[a4paper]{article}

\usepackage{INTERSPEECH2020}

\usepackage{multirow,comment}
\title{Improving Noise Robustness In Speaker Recognition Using A Two-Stage Attention Model}
\name{Yanpei Shi, Qiang Huang, Thomas Hain}
\address{ Speech and Hearing Research Group\\
          Department of Computer Science, University of Sheffield}
\email{\{YShi30, qiang.huang, t.hain\}@sheffield.ac.uk}

\begin{document}

\maketitle
\begin{abstract}
While the use of deep neural networks has significantly boosted speaker recognition performance,
it is still challenging to separate speakers in poor acoustic environments. 
To improve robustness of speaker recognition system performance in noise, a novel two-stage attention mechanism which can be used in existing 
architectures such as Time Delay Neural Networks (TDNNs) and Convolutional Neural Networks (CNNs) is proposed.
Noise is known to often mask important information in both 
time and frequency domain. The proposed mechanism allows the models to concentrate on 
reliable time/frequency components of the signal. 
The proposed approach is evaluated using the Voxceleb1 dataset, which aims
at assessment of speaker recognition in real world situations. In 
addition three types of noise at different signal-noise-ratios (SNRs) were added
for this work. The proposed mechanism is compared with three strong baselines:
X-vectors, Attentive X-vector, and Resnet-34. Results on both identification 
and verification tasks show that the two-stage attention mechanism
consistently improves upon these for all noise conditions. 
%

\end{abstract}
\noindent\textbf{Index Terms}: Robust Speaker Recognition, Attention Mechanism, Time-Delay Neural Network, Convolutional Neural Network, Two-Stage Attention. 

\section{Introduction}\label{introduction}
The goal of speaker recognition is to recognize a speaker from the characteristics of 
voices \cite{poddar2017speaker}. 
I-vector \cite{dehak2010front} based on GMM-UBM
was developed and widely used to extract acoustic features for speaker recognition.
Speech signals in real environment are often corrupted by different types of background noise \cite{ming2007robust}.
This might influence some key acoustic features of speakers 
and thus make speaker recognition in noise conditions a challenging task.

In recent years, recognizing speaker identities from audio signal using deep neural networks 
has been an active research area and different speaker modelling approaches \cite{poddar2017speaker, snyder2018x, variani2014deep} 
were proposed. Variani, et al. developed the d-vector which uses multiple fully-connected neural network layers \cite{variani2014deep}.
In \cite{snyder2018x}, Snyder, et al. proposed X-vectors, which consists of a TDNN structure that can
model relationships in wide temporal contexts and computes speaker embeddings from variable length
acoustic segments.

To further tackle interferences caused by background noise, an attention mechanism \cite{hu2019introductory} 
was used to allocate weights on different part of data and highlight the information
which is relevant to targets. 
For speaker recognition, there are some previous studies
that use attention models in time dimension \cite{zhu2018self,okabe2018attentive,wang2018attention, rahman2018attention}. 
Wang, et al. \cite{wang2018attention} used an attentive X-vector where a self-attention layer was added 
before a statistics pooling layer to weight each frame. 
Rahman, et al. \cite{rahman2018attention} jointly used attention model and K-max pooling to
selects the most relevant features. 

In addition to speaker recognition, the attention model has also been widely used in  
natural language processing \cite{bahdanau2014neural, luong2015effective, yang2016hierarchical, huang2016attention}, 
speech recognition \cite{moritz2019triggered, mirsamadi2017automatic,zhang2018attention, chorowski2015attention}, 
and computer vision \cite{xu2015show, li2017image, woo2018cbam, wang2017residual, oktay2018attention, mejjati2018unsupervised}.
To further improve the robustness of the attention model, some previous studies used two attention models within one framework.
Luong, et al. \cite{luong2015effective} used global attention and local attention, where 
global attention attends to the whole input sentence and local 
attention only looks at a part of the input sentence.
Li, et al. \cite{li2017image} applied global and local attention in image processing to further improve the performance. 
Woo, et al. \cite{woo2018cbam} used spatial attention and channel attention to extract salient features from input data.

To mitigate the interferences caused by noise, this work proposes a two-stage attention model. 
The two-stage attention simulates the procedure of designing a noise filter.
To better reduce the pass-band ripples and the transition band,
a good-quality filter is generally designed by cascading several lower-order filters 
instead of directly building a high-order filters  \cite{porle2015survey,imai1988design}.
Inspired by this case, the two attention modules in this work are used sequentially to process
features in time and frequency domain, which is like cascading two filters.  
This might be able to reduce some possible impacts caused by over-fitting when training models
on noise corrupted time-frequency features.
In the two-stage attention framework, the first attention module works
on elements of feature vectors and is called as ``frequency attention model''. The second one
computes weights on data frames and is called as
``time attention model''. For comparison, the case of running the
two attention models in parallel is also introduced in the following section.

The rest of the paper is organized as follow: Section \ref{{Model Architecture}} presents the model architectures of our approaches. 
Section \ref{Experiment} depicts the data we use, experimental setup, and the baselines to be compared.
We show the obtained results in Section \ref{Results}, and finally draw a conclusion in Section \ref{conclusion}.

\section{Model Architecture}\label{{Model Architecture}}

\begin{figure}[t]
	\centering
	\includegraphics[height=7.5cm,width=8cm]{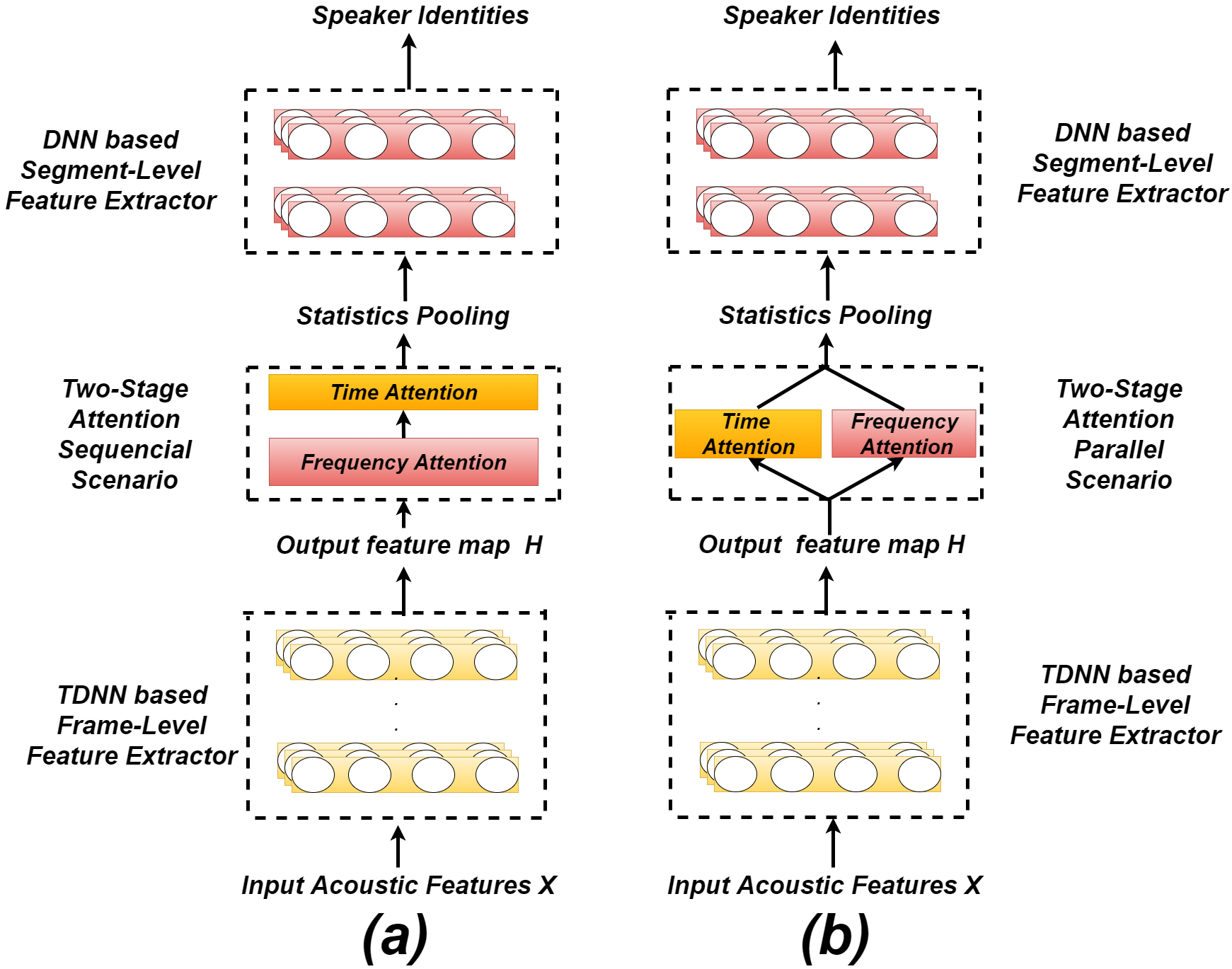}
	\caption{Architecture of the proposed models, (a): cascade of the two stage attention model and (b): parallel of the two stage attention model}
	\label{fig:TS_attention}
\end{figure}

Figure \ref{fig:TS_attention} shows the architectures of our approaches,
implementing the two attention models in cascade (a) and in parallel (b).
From the input to output, each sub-figure
consists of a time delay neural network (TDNN),
a two-stage attention model, a statistics pooling layer, and two fully connected layers. 
The details of the TDNN model could be found in \cite{snyder2018x}.

The input data is $\boldsymbol X = \{\boldsymbol x_{1},\boldsymbol x_{2},...,\boldsymbol x_{T}\}$ 
($\boldsymbol X \in  \mathcal {R}^{T \times L})$, 
where $T$ represents the sequence length, $L$ represents the dimension of each feature vector, and $x_{i}$ denotes
the $i$th acoustic feature vector extracted from the speech signal. The TDNN operates as a frame-level feature extractor
and $\boldsymbol H = \{\boldsymbol h_{1},\boldsymbol h_{2},...,\boldsymbol h_{T}\}$ 
($\boldsymbol H \in \mathcal {R}^{T \times F}$) denotes its output,
where $T$ is its length (same as $X$), $F$ is the frequency dimension, and
$h_{i}$ denotes the $i$th vector for each frame \cite{ming2007robust}.
The two-stage attention module consists of a time attention model and a frequency attention model,
whose input is $\boldsymbol H$ and output is denoted by $\boldsymbol H^{''} = \{\boldsymbol h^{''}_{1},\boldsymbol h^{''}_{2},...,\boldsymbol h^{''}_{T}\}$,
where $\boldsymbol H^{''}$ has the same dimension as $\boldsymbol H$.

\subsection{Two-stage Attention Model}\label{Two-stage Attention Approach}
\subsubsection{Cascade Attention Model}
\begin{figure}[h]
	\centering
	\includegraphics[height=8cm,width=7.5cm]{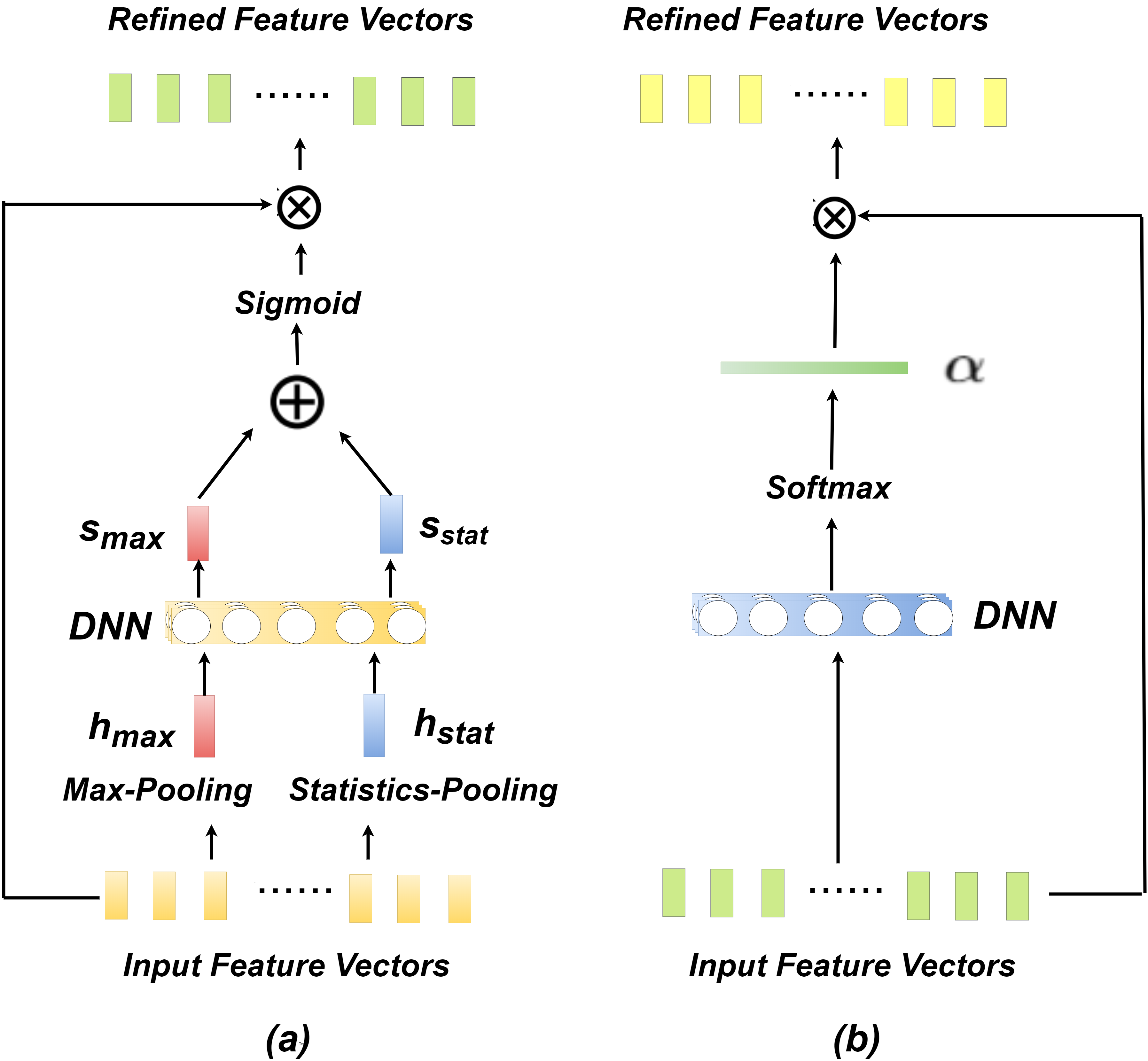}
	\caption{Architecture of the two-stage attention model: the frequency attention model (a) and the time attention model (b)}
	\label{two-stage-attention}
\end{figure}

As shown in Figure \ref{fig:TS_attention} (a), the two-stage attention models are applied sequentially, where
the frequency attention model is followed by the time attention model.

The frequency attention model uses a self-attention structure to allocate
weights for each element on the frequency dimension of $\boldsymbol H$. 
In Eq \ref{equ:EAM},  $\boldsymbol F'_{freq}$ is obtained by copying the frequency attention weight vector 
$\boldsymbol F_{freq}(\boldsymbol H) \in \mathcal {R}^{1 \times F}$ along the temporal dimension.
Element-wise multiplication ($\odot$) of $\boldsymbol F'_{freq}$ by $\boldsymbol H$ results in the output of frequency attention $\boldsymbol H^{'} \in \mathcal {R}^{T \times F}$:
\begin{equation}\label{equ:EAM}
\boldsymbol H^{'} = \boldsymbol F'_{freq} \odot \boldsymbol H
\end{equation}

Figure \ref{two-stage-attention} (a) shows the computation of $\boldsymbol F_{freq}(\boldsymbol H)$, which is defined as:
\begin{equation}\label{feature wise attention}
\begin{aligned}
\boldsymbol F_{freq}(\boldsymbol H) &= Sigmoid(\boldsymbol s_{stat}+\boldsymbol s_{max})\\
\boldsymbol s_{stat} &= Relu(\boldsymbol h_{stat} \boldsymbol W^{c}_{0}+\boldsymbol b^{c}_{0})\boldsymbol W^{c}_{1}\\
\boldsymbol s_{max} &= Relu(\boldsymbol h_{max}\boldsymbol W^{c}_{0}+\boldsymbol b^{c}_{0})\boldsymbol W^{c}_{1}
\end{aligned}
\end{equation}
The frequency attention model employs two different pooling mechanisms, max-pooling and statistics-pooling.
The output of max-pooling $\boldsymbol h_{max} \in \mathcal {R}^{1 \times F}$ is used to compute 
$\boldsymbol s_{max} \in \mathcal {R}^{1 \times F}$ after
employing a linear mapping and an activation function ($Relu$ \cite{sibi2013analysis}). 
The statistics-pooling outputs are $\boldsymbol h_{avg} \in \mathcal {R}^{1 \times F}$ and $\boldsymbol h_{std} \in \mathcal {R}^{1 \times F}$. They are then summed together into $\boldsymbol h_{stat} \in \mathcal {R}^{1 \times F}$
The details on how to implement this type of pooling is referred in \cite{wang2018attention}. 

$\boldsymbol W^{c}_{0} \in \mathcal {R}^{F \times K}$, $\boldsymbol b^{c}_{0} \in \mathcal {R}^{1 \times K}$ and $\boldsymbol W^{c}_{1} \in \mathcal {R}^{K \times F}$ shown in Eq \ref{feature wise attention} are the parameters of the frequency attention. The parameter K is used to control the number of parameters in the frequency attention model, and it is set to 100 in this work. 
The weights are finally obtained using a sigmoid function \cite{sibi2013analysis} on the sum of $\boldsymbol s_{max}$ and $\boldsymbol s_{stat}$.

The time attention model also uses a self-attention structure whose input is $\boldsymbol H^{'}$ and output is $\boldsymbol H^{''}$: 
\begin{equation}\label{equ:FAM}
\boldsymbol H^{''} = \boldsymbol F'_{time} \odot \boldsymbol H^{'}, 
\end{equation}
where $\boldsymbol F'_{time}$ is obtained by copying
the time attention weight vector $\boldsymbol F_{time}(\boldsymbol H^{'}) \in \mathcal {R}^{T \times 1}$ 
along frequency dimension. 

$\boldsymbol F_{time}(\boldsymbol H^{'})$ is defined as:
\begin{equation}\label{equ:f_frame}
\boldsymbol F_{time}(\boldsymbol H^{'}) = \boldsymbol \alpha ,
\end{equation}
where $\boldsymbol \alpha \in \mathcal {R}^{T \times 1}$ is a normalized score vector $\boldsymbol \alpha = \{\alpha_{1}, \alpha_{2}, ... , \alpha_{T} \}$, 
$\alpha_{t}$ denotes the scalar weight for each frame and is computed using a Softmax function \cite{wang2018attention,zhu2018self}:
\begin{equation}\label{att}
\begin{aligned}
\alpha_{t} &= \frac{exp(s_{t})}{\sum_{i=0}^{T}exp(s_{t})} \\
s_{t} &= Relu(\boldsymbol h^{'}_{t} \boldsymbol W_{0}+\boldsymbol b_{0})\boldsymbol W_{1}
\end{aligned}
\end{equation}
$\boldsymbol W_{0} \in \mathcal {R}^{F \times F}$, $\boldsymbol b_{0} \in \mathcal {R}^{1 \times F}$ and $\boldsymbol W_{1} \in \mathcal {R}^{F \times 1}$ are the parameters used in time attention model.

\subsubsection{Parallel Attention Model}
As shown in figure \ref{fig:TS_attention} (b), when running the two stage attention models in parallel, $\boldsymbol F_{time}$ and $\boldsymbol F_{freq}$ share the
same input $\boldsymbol H$. Their outputs are merged by firstly broadcast into the same dimension and added together, and then multiplying $\boldsymbol H$ element wisely ($\odot$):
\begin{equation}\label{parallel}
\boldsymbol H^{''} = (\gamma \cdot \boldsymbol F_{freq}(\boldsymbol H) + (1-\gamma) \cdot \boldsymbol F_{time}(\boldsymbol H)) \odot \boldsymbol H ,
\end{equation}
where $\boldsymbol F_{time}(\boldsymbol H)$ and $\boldsymbol F_{freq}\boldsymbol (\boldsymbol H)$ are computed using Eq \ref{feature wise attention} and Eq \ref{equ:f_frame}, respectively. $\gamma$ is a hyper-parameter of the parallel two-stage attention model.

\subsection{Two-stage attention in CNN architecture}
In addition to the TDNN based model depicted in the last subsection, the two-stage attention module
can be also applied in CNN based architecture, such as Resnet-34 \cite{zeinali2019but}. 

Suppose $\boldsymbol H_{k} \in \mathcal {R}^{T_{k} \times F_{k} \times C_{k}}$ is the output feature map of the $k$th residual block in a Resnet-34 model, 
where $T_{k}$, $F_{k}$, $C_{k}$ represents the time, frequency and feature dimension. 
Then, $\boldsymbol H_{k}$ is reshaped into $\boldsymbol H_{k}^{'} \in \mathcal {R}^{T_{k} \times F_{k}C_{k}} $, where the frequency and feature dimension are combined \cite{miao2019new}. 
Similar to that in TDNN based two-stage attention model, frequency and time attention are applied on $\boldsymbol H_{k}^{'}$.
Both the frequency and time attention in CNN architecture use Eq \ref{feature wise attention} to compute the corresponding attention weights, 
as the time dimension in CNN architecture is compressed and the use of Softmax function might lose information \cite{woo2018cbam}.

Two-stage attention is applied at the end of each residual block in the Resnet-34 architecture, instead of using it once in TDNN based architecture.

\vspace*{-2mm}
\section{Experiments}\label{Experiment}
\vspace*{-2mm}
\begin{table*}[t]
\renewcommand{\multirowsetup}{\centering}  
\renewcommand\arraystretch{1.0}
\setlength{\tabcolsep}{1.8mm}
\centering  
\footnotesize
\begin{tabular}{c|c|c|c|c|c|c|c|c|c|c|c}
\hline
\multirow{2}{*}{\textbf{Noise Type}}& \multirow{2}{*}{\textbf{SNR}} & \multicolumn{2}{c|}{\textbf{TDNN}}&\multicolumn{2}{c|}{\textbf{TDNN+ATT}} &  \multicolumn{2}{c|}{\textbf{TF}}&\multicolumn{2}{c|}{\textbf{FT}}&\multicolumn{2}{c}{\textbf{Para}}\\

\cline{3-12}
& & Top1 (\%)& EER (\%)&Top1 (\%)& EER (\%)& Top1 (\%)& EER (\%)&Top1 (\%)& EER (\%)&Top1 (\%)& EER (\%)\\

\hline
\multirow{5}{*}{\textbf{Noise}}&
0 &74.6 & 12.26& 75.8& 11.32& 76.0  &11.13  & \textbf{77.2}&\textbf{10.68} &76.8 & 10.92\\
&5&79.5 & 10.01& 79.4&9.26 & 80.0 & 9.02 & \textbf{81.3}&\textbf{8.82} &80.8 &9.04\\
&10& 83.1& 8.33& 84.0& 7.77& 84.6 & 7.42 & \textbf{86.6}& \textbf{7.04}& 86.3& 7.32\\
&15&85.0 & 7.25& 86.3& 6.76& 86.9 & 6.55 &\textbf{88.3} & \textbf{6.25}&87.8 & 6.40\\
&20&87.9 &6.91 & 87.8& 6.02& 88.2 & 5.99 & 89.6&\textbf{5.84}& \textbf{89.8}& 5.88\\
\hline

\multirow{5}{*}{\textbf{Music}}&
0 &68.2 & 14.15&70.1 & 12.92& 71.2 & 12.69 & \textbf{73.4}& \textbf{12.48}& 72.6& 12.64\\
&5& 72.0&11.03 & 73.5& 10.04& 74.0 & 9.89 &\textbf{75.9} &\textbf{9.34} & 75.1&9.52\\
&10& 79.4& 9.35& 81.0& 8.64& 82.1 & 8.28 &\textbf{84.0} &\textbf{8.17} & 82.8&8.35\\
&15& 84.2& 8.41& 86.6& 8.08& 86.8 &7.70 & \textbf{87.1}& 7.33& 86.8&\textbf{7.29}\\
&20& 86.1& 6.79& 88.0& 6.25& 88.5 & 6.17 &\textbf{89.3} & 6.04& 89.0&\textbf{6.01}\\
\hline

\multirow{5}{*}{\textbf{Babble}}&
0 &64.1 & 30.02& 65.2& 27.77& 67.1 & 27.27 & \textbf{68.9}& \textbf{26.53}& 67.6&26.94\\
&5& 70.5& 16.46& 71.4& 15.32& 73.0 & 15.05 & \textbf{75.0}& \textbf{14.22}& 73.8&14.83\\
&10&77.4 & 13.26& 77.0& 12.53& 78.5 & 12.44 & \textbf{79.8}& \textbf{12.13}&78.8 &12.30\\
&15& 83.5& 9.10& 84.5& 8.31& 86.0 & 8.06 & \textbf{87.1}& \textbf{7.99}& 86.9&8.11\\
&20& 86.6& 7.95& 86.9& 7.22& 87.9 & 7.04 & \textbf{88.6}& \textbf{6.74}& 88.2&6.91\\
\hline

\textbf{Original}&  & 88.2&5.47 &89.2  &5.06  & 89.9 & 5.01 & \textbf{91.1} & \textbf{4.91}& 90.7& 4.99\\
\hline
\end{tabular}

\caption{Speaker identification and verification results for different noise types (Noise, Music and Babble) at different SNR (0-20 dB), and the original Voxceleb1 test set. Five different models are tested: X-vector, Attentive X-vector, two-stage attention with time attention first, frequency attention first and parallel. $\gamma$ is set to 0.5 in parallel scenario.}
\label{tdnn_results}
\end{table*}

\begin{table*}[t]
\renewcommand{\multirowsetup}{\centering}  
\renewcommand\arraystretch{1.0}
\setlength{\tabcolsep}{2mm}
\centering  
\footnotesize
\begin{tabular}{c|c|c|c|c|c|c|c|c|c|c|c}
\hline
\multirow{2}{*}{\textbf{Noise Type}}& \multirow{2}{*}{\textbf{SNR}} & \multicolumn{2}{c|}{\textbf{Resnet34}}&\multicolumn{2}{c|}{\textbf{T}}&\multicolumn{2}{c|}{\textbf{TF}} &  \multicolumn{2}{c|}{\textbf{FT}}&\multicolumn{2}{c}{\textbf{Para}}\\

\cline{3-12}
& & Top1 (\%)& EER (\%)&Top1 (\%)& EER (\%)& Top1 (\%)& EER (\%)&Top1 (\%)& EER (\%)&Top1 (\%)& EER (\%)\\

\hline
\multirow{5}{*}{\textbf{Noise}}&
0 &77.3 & 10.03& 77.7 & 9.94& 78.0 & 9.81& \textbf{79.4} & \textbf{9.58} & 78.8& 9.72\\
&5& 81.5& 8.10& 82.0 & 8.03&82.8& 7.97& \textbf{84.6} & \textbf{7.68} & 83.2& 7.81\\
&10& 82.4& 6.92& 82.9 &6.76 &83.1& 6.65& \textbf{86.5} & \textbf{6.26} & 84.9& 6.57\\
&15& 84.4& 6.45& 85.1& 6.37& 85.9& 6.33& \textbf{87.2} & \textbf{5.99} & 86.3& 6.13\\
&20& 87.2& 5.72& 87.9 & 5.60 &88.6& 5.58& \textbf{89.8} & 5.43 & 89.2& \textbf{5.41}\\
\hline

\multirow{5}{*}{\textbf{Music}}&
0 & 72.5& 12.16& 73.0 & 12.04&73.4& 11.89& \textbf{75.6} & \textbf{11.68} & 74.4& 11.79\\
&5& 76.9& 9.28& 77.4 & 9.09& 77.4&  9.01& \textbf{78.4}& \textbf{8.69} & 77.8& 8.85\\
&10& 83.8& 8.25& 84.6 & 8.17& 85.1&  8.11 & \textbf{86.8}& \textbf{7.93} & 86.5& 8.03\\
&15& 86.1& 7.34& 87.0 & 7.19&87.6&  7.12 &88.3 & \textbf{7.01} & \textbf{88.5}& 7.09\\
&20& 87.4& 6.39& 88.2 & 6.28&88.7&  6.20 &\textbf{89.7} & 5.92 & 89.2& 6.07\\
\hline

\multirow{5}{*}{\textbf{Babble}}&
0 & 69.3& 28.95& 69.7 & 28.60&70.2& 28.17& \textbf{72.5} & \textbf{27.79} & 71.7& 28.02\\
&5& 76.2& 17.36& 76.9 & 17.04&77.1&16.93 & \textbf{78.3} & \textbf{16.17} & 77.7& 16.59\\
&10& 81.4&12.04& 81.5 & 11.78&81.9& 11.59& \textbf{83.2} &  \textbf{10.82}& 82.4& 11.35\\
&15& 84.0& 8.96& 84.3 & 8.88&84.4 &8.86& \textbf{86.0} & \textbf{8.79} & 85.1& 8.83\\
&20& 87.8& 7.05& 88.2 & 7.01&88.0& 6.98& \textbf{88.7} & \textbf{6.72} & 88.5& 6.93\\
\hline

\textbf{Original}&  & 90.0&5.35 & 90.3& 5.04&90.6 & 4.98& \textbf{92.0} & \textbf{4.81} &91.2 & 4.90\\
\hline
\end{tabular}

\caption{Speaker identification and verification results for different noise types (Noise, Music and Babble) at different SNR (0-20 dB), and the original Voxceleb1 test set. Five different models are tested: Resnet-34, Resnet-34 with time attention, two-stage attention with time attention first, frequency attention first and parallel. $\gamma$ is set to 0.5 in parallel scenario.}
\label{cnn_results}
\end{table*}

\begin{table}[t]
\renewcommand{\multirowsetup}{\centering}  
\renewcommand\arraystretch{1.0}
\setlength{\tabcolsep}{3mm}
\centering  
\footnotesize
\begin{tabular}{c|c|c|c|c}
\hline
& \multicolumn{2}{c|}{\textbf{TDNN+Para}}&\multicolumn{2}{c}{\textbf{CNN+Para}}\\

\cline{1-5}
$\boldsymbol \gamma$& Top1 (\%)& EER (\%)&Top1 (\%)& EER (\%)\\

\hline
0.0&65.2 & 27.77& 69.7 & 28.60\\
0.2&66.0 & 27.28& 70.4 & 28.34\\
0.4&67.1 & 27.01& 70.8 & 28.13\\
0.6&\textbf{67.9} & \textbf{26.59}& 71.3 & 27.88\\
0.8&67.5 & 26.89& \textbf{72.2} & \textbf{27.73}\\
1.0&66.9 & 27.04& 71.4 & 28.02\\

\hline
\end{tabular}

\caption{Speaker identification and verification results on TDNN+Para and CNN+Para with different $\gamma$ value, the noise type is Babble and SNR level is 0 dB.}
\label{gamma}
\end{table}

%
\subsection{Data}\label{Data}
In this work, the Voxceleb1 \cite{nagrani2017voxceleb} dataset is employed
as it is one of the most widely used datasets for speaker identification and verification.
This dataset is extracted from Youtube videos, collected "in the wild",
and has an official train-test split for both speaker identification and verification tasks. 
For the speaker identification task, the training set and test set contains the same number of speakers. For the speaker verification split, the test set contain 37720 test pairs, 40 distinct speakers totally.
In order to test the robustness of the proposed model in different noise conditions, 
the MUSAN dataset \cite{snyder2015musan} is used to generate noise corrupted signals by mixing with the utterances from Voxceleb1.  
The MUSAN dataset contains the recordings of three noise types: general noise, music and speech. 

In experiments, two kinds of features are extracted after speech streams are segmented by a 25-ms sliding window with a 10-ms hop.
For the TDNN based models, 40-dimensional log-Mel filter-bank vectors are used.
For the CNN based modesl, after using a 512-point FFT on speech segments, 257-dimension spectrograms, including a DC component, are used as input features.

\subsection{Experiment Setup}

In this work, both speaker identification and speaker verification tasks are conducted to test the proposed model in close-set and open-set speaker recognition \cite{nagrani2017voxceleb}.
For both two tasks, the training set is augmented by mixing Voxceleb1 data with noise signals at a random SNR level (0, 5, 10, 15 and 20 dB). 
The test utterances are mixed with a certain kind of noise with one of the five SNR levels (0, 5, 10, 15 and 20 dB). 

For the speaker identification task, the models are trained using normal softmax function with cross-entropy loss.The top-1 accuracy is used as the evaluation metric \cite{ge2017neural}. In the speaker verification task, the models are trained using normal softmax function with cross-entropy loss, and then fine-tuned using AM-Softmax loss (m is set to 0.3, s is set to 35) \cite{wang2018additive}. 
Equal Error Rate (EER) is used as the evaluation metric \cite{cheng2004method}.

The experiments are conducted using TDNN or CNN based models respectively. For the TDNN based models, 
X-vectors \cite{snyder2018x} and attentive X-vectors \cite{zhu2018self,okabe2018attentive,wang2018attention, rahman2018attention} are used as the baseline methods. Three combinations of time and frequency attention scenarios are tested: time attention first (T-F), frequency attention first (F-T) and parallel (Para).
For the CNN based model, Resnet-34 \cite{zeinali2019but} is employed. Three different attention scenarios (F-T, T-F and Para) are applied on Resnet-34 architecture. The scenario that used a time attention (T) only is also tested.

\subsection{Parameter Configuration}
\vspace*{-2mm}
In this work, the Adam optimiser \cite{Kingma2014AdamAM} is used in training. The initial learning rate is 1e-4 with decay rate 0.95 for each epoch. For the use of AM-Softmax loss in speaker verification task, the initial learning rate is set to 5e-5.

\vspace*{-3mm}
\section{Results}\label{Results}
\vspace*{-2mm}

Table \ref{tdnn_results} and \ref{cnn_results} show the speaker identification and verification results obtained using the TDNN based models and CNN based models. It is clear that the proposed two-stage attention model performs better than the baseline methods (X-vector, Attentive X-vector and Resnet-34) in almost all noise conditions at different SNR levels.

When the noise type becomes more complex and the noise level becomes larger, such as babble and music noise type at 0 and 5 dB, the gap between the proposed two-stage attention model and the corresponding baselines becomes larger. In Babble noise type at 0 dB, the F-T scenario (frequency first two-stage attention) in TDNN model reaches more than 3\% relatively improvement on speaker identification and verification task. The same improvement also achieved by the F-T on CNN model.

Compared with using time attention only (Attentive X-vector and Resnet-34+time attention), the combination of time and frequency attention obtained better results. It is clear that the use of the two attention model is better than the use of a single attention model running only on time dimension. Multiple attention models
enable the system to learn more information relevant to target speakers than the baseline 
by highlighting the important features in both time and frequency dimension.

Comparing with different combination strategy of two-stage attention, frequency attention first (F-T) performs better than other models in most noise conditions and levels. On TDNN based models, F-T reaches 91.1\% speaker identification accuracy and 4.91\% equal error rate on speaker verification task on the original Voxceleb1 test set. In CNN architecture, F-T reaches 92.0\% accuracy in speaker identification task and 4.81\% equal error rate in speaker verification task. 

The reason why frequency attention first obtained better results than Parallel scenario might because that 
the use of the cascade model is actually similar to the design of a digital filter as mentioned
in Section \ref{introduction}, by which it is probably able to
provide some good constraints in information selection and model optimisation \cite{porle2015survey,imai1988design}.    
Compared to the cascade model, the two attention models running in parallel might work 
a bit more independently, resulting in less constraints to the data to be processed. 

The reason why T-F scenario performs worse than F-T scenario might because in noise conditions, the frequency dimension contains more information than time dimension, and T-F scenario applied time attention first, which might allocate lower weight to some frames that lose some important information.

To further test and compare the performance of parallel scenario (Para), the weight value $\gamma$ in para scenario is tuned from 0 to 1. The different $\gamma$ value is tested on Babble noise type and SNR value is equals to 0 dB. 
Table \ref{gamma} shows the speaker identification accuracies on Para scenario with different $\gamma$ value in TDNN and CNN based models. Results show that with the increase of $\gamma$, the accuracies become higher and the equal error rate lower down. For TDNN model, it reaches a peak when $\gamma$ is equals to 0.6, and when  $\gamma$ =0.8, CNN model reaches its peak. This results shows that the frequency attention in both TDNN and CNN architectures contributes more to the recognition results, it also shows a possible reason why the results of time attention first (T-F) is worse than that of frequency attention first (F-T). 

\vspace*{-2mm}
\section{Conclusion and Future Work}\label{conclusion}
\vspace*{-2mm}
In this paper a two-stage attention model was proposed to recognize speakers in noise environment. 
The proposed model contains a frequency attention model and a time attention model. The two attention model can be either applied sequentially or in parallel, and the combination can be used in the current widely used speaker recognition models. The speaker identification and verification results in different noise conditions and levels on Voxceleb1 dataset show strong robustness against the effect caused by noise.
In future work, the developed approaches will be tested on more datasets, such as
Voxceleb2 for speak recognition.
Moreover, some complex network architectures and noise types will be investigated.

\begin{center}
\large{\textbf{Acknowledgement}}
\end{center}
\vspace*{-2mm}
This work was in part supported by Innovate UK Grant number 104264 MAUDIE.

\bibliographystyle{IEEEtran}

\bibliography{mybib}


\end{document}